\documentclass[aps,showpacs,preprintnumbers,nofootinbib,tighten]{revtex4}
\usepackage{graphicx}
\usepackage{dcolumn}
\usepackage{bm}
\usepackage{amssymb,amsmath}
\usepackage{epsfig}   

\input epsf

\begin{document}

\title{\Large{Field Theoretic Description of Ultrarelativistic Electron-Positron Plasmas}}

\affiliation{Markus H. Thoma}

\address{Max-Planck-Institut f\"ur extraterrestrische Physik,
P.O. Box 1312, 85741 Garching, Germany}


\vspace{0.4in}

\begin{abstract}
Ultrarelativistic electron-positron plasmas can be produced in high-intensity laser fields and play a role 
in various astrophysical situations. Their properties can be calculated using QED at
finite temperature. Here we will use perturbative QED at finite temperature for calculating
various important properties, such as
the equation of state, dispersion relations of collective plasma modes of photons and
electrons, Debye screening, damping rates, mean free paths, collision times, transport coefficients,
and particle production rates, of ultrarelativistic electron-positron plasmas. In particular,
we will focus on electron-positron plasmas produced with ultra-strong lasers.  
\end{abstract}

\maketitle

\vspace{0.2in}

\tableofcontents

\section{Introduction}

Ultrarelativistic plasmas, in which the thermal energy of the particles
is much larger than their rest mass energy, were discussed first in the fifties
in the context of astrophysics. They occur in the early Universe as well as stellar 
and galactic high-energy processes (Zel'dovich and Novikov, 1971; Zel'dovich and Novikov, 
1983; Raffelt, 1996). For example, ultrarelativistic electron-positron plasmas (EEP) 
can be created either by high temperatures in supernovae explosions (see e.g.
Hardy and Thoma (2001)) or strong magnetic fields
in so-called magnetars (Beskin {\it et al.}, 1993). The theoretical description
of ultrarelativistic plasmas is based on transport theory (Silin, 1960) or
thermal field theory (Tsytovich, 1961). The latter is an extension of quantum
field theory, such as QED or QCD, in vacuum to finite temperature (or chemical
potential) describing high-energy particle interactions in matter. For this purpose,
two different approaches were developed, the imaginary time formalism (Matsubara,
1955; Kapusta, 1989) and the real time formalism (Landsmann and van Weert, 1987).

In the eighties a new type of ultrarelativistic plasmas became of interest, the 
quark-gluon plasma (QGP) (M\"uller, 1985). It corresponds to a new state of matter of deconfined
quarks and gluons at extremely high temperatures above $k_BT=150$ MeV. The early Universe
should have been in this state for the first few microseconds after the Big Bang. Furthermore, the QGP
is expected to be created in relativistic heavy-ion collisions, in particular 
within the accelerator experiment RHIC (Gyulassy and McLerran, 2005). There a hot
and dense fireball (``little bang'') of the size of an atomic nucleus is produced, 
which could be in the QGP phase for less than $10^{-22}$ s. Therefore the
QGP in those experiments cannot be observed directly but its discovery relies on the 
comparison of theoretically predicted signatures with experimental data (``circumstantial
evidence''). For understanding the properties of the QGP and for calculating signatures
for its formation, QCD at finite temperature has been applied. Besides a non-perturbative
method based on a numerical solution of the QCD equations (lattice QCD), perturbation
theory at finite temperature has been used widely (Thoma, 1995a; Le Bellac, 1996). 
Lattice QCD allows only the calculation of static quantities, and hence not of most of the
proposed signatures following from particle production. Perturbative QCD, on the other
hand, has been used for computing static as well as dynamic quantities. However, its
predictions are limited  by the fact that the QGP is a strongly-coupled plasma
at temperatures achievable in the experiments. (At extremely high
temperatures far above the transition temperature perturbative QCD should be 
reliable due to asymptotic freedom of QCD.) Nonetheless, a wide variety of 
properties and quantities of the QGP have been considered in this way. It started
with the calculation of self-energies and dispersion relations in the high-temperature
limit (Klimov, 1982; Weldon, 1982a; Weldon, 1982b). In the high-temperature limit
the results differ from the QED results only by trivial factors containing the QCD
degrees of freedom (color, flavor). In addition, it can be shown that the gluon
self-energy or polarization tensor, which is directly related to the dielectric
function (see e.g. Elze and Heinz (1989)), can be derived also 
within the transport (Vlasov) approach (Silin, 1960), since the high-temperature limit
corresponds to the classical limit. Other quantities, however, such as damping or
production rates and transport coefficients, require a quantum field theoretic
treatment. In particular, a resummation of certain diagrams, the hard thermal loops (HTL),
is needed for a consistent description, i.e. to obtain gauge invariant, infrared finite (or
improved) results complete to leading order in the coupling constant (Braaten and Pisarski, 1990).
Details about quantum field theoretic methods and their application to
the physics of the QGP can be found in the review article by Thoma (1995a). 
             
Recently a third possibility to study ultrarelativistic plasmas has been suggested: in extremely 
strong laser fields the creation of an ultrarelativistic EPP with temperatures around
10 MeV could be realized soon (Liang {\it et al.}, 1998). For example two opposite, circular 
polarized laser pulses with a duration of 330 fs  
and an intensity of $7\times 10^{21}$ W/cm$^2$ could hit a thin gold foil. In this way the target
electrons can be heated up to about 10 MeV producing an ultrarelativistic EPP by pair creation. The 
positron density could be about $5\times 10^{22}$ cm$^{-3}$ (Shen and Meyer-ter-Vehn, 2001). This will allow for the first 
time to study an ultrarelativistic EPP in the laboratory. Therefore predictions of the physical properties 
of such a system are needed.

The interaction of relativistic electrons and positrons is described by QED. In the presence of a thermal 
plasma background (heat bath) QED at finite temperature has to be considered. As discussed above, perturbative 
methods based on the imaginary (Matsubara) or real time formalism have been developed and applied
to the physics of a QGP. Here we will transfer the methods and results 
for properties of a QGP to the case of an EPP, where a perturbative treatment is more reliable than in 
the case of a strongly coupled QGP. Some applications of thermal field theory to astrophysical plasmas 
have been discussed before (see e.g. Thoma (2002) and references therein and Altherr and Kraemmer (1992)). 

An important difference to non-relativistic ion-electron plasmas (Lifshitz and Pitaevskii, 1981)
are the relevant scales. In the non-relativistic case 
case they are given by the masses of the plasma particles and the temperature $T$. For example, the 
electron plasma frequency reads
\begin{equation}
\omega_{pl}=\sqrt{\frac{4\pi e^2\rho_e}{m_e}}
\label{e1}
\end{equation}     
and the Debye screening length due to the electrons in the plasma
\begin{equation}
\lambda_D=\sqrt{\frac{k_BT_e}{4\pi e^2\rho_e}},
\label{e2}
\end{equation}     
where $\rho_e$ is the electron number density, $T_e$ the temperature of the electron component,  
and $m_e$ the electron mass.

In an ultrarelativistic plasma with $T \gg m$ the masses can be neglected and the important scales are
the temperature $T$, called the hard scale, and the soft scale $eT$, which determines the collective 
phenomena as we will see below. Here we use natural units, i.e. $\hbar =c=k_B=1$, as usual in quantum 
field theory. In these units $e=0.3$ corresponding to a fine structure constant $\alpha =e^2/(4\pi)=1/137$.
For converting to SI-units we use $1\> {\rm MeV}\> \hat =\> 1.60\times 10^{-13}\> {\rm J}
\> \hat =\> 5.08\times 10^{12}\> {\rm m}^{-1}\> \hat =\> 1.52\times 10^{21}\> {\rm s}^{-1}$.

In the next section we will discuss the equation of state of an equilibrated ultrarelativistic EPP.
Then collective phenomena will be considered. Afterwards transport properties and particle
production will be discussed. Finally we will describe properties of an EPP which is not in
chemical equilibrium as in the case of laser induced plasmas. We will not consider here the 
formation process and equilibration of an EPP.  
 
Many results presented here can be found in the literature for the case of a QGP (Thoma, 1995a)
differing only by numerical 
factors, e.g. number of degrees of freedom. The purpose of this colloquium is to summarize these results and
to extend them to laser induced EPPs as a reference for future experiments. As an example we consider
a temperature of 10 MeV as it can be typically realized in laser produced and supernovae EPPs.

Laser produced QED plasmas have also been discussed recently in two review articles (Mourou {\it et al.}, 2006;
Marklund and Shukla, 2006) with emphasis on the production mechanism and non-linear effects. Here, however, we want to
focus on the properties of an equilibrated EPP as they can be calculated from perturbative QED. Such an
EPP in thermal and chemical equlibrium might be the outcome of future laser experiments if the intensity
can be increased further on. As an example, we have chosen the predictions of Shen and Meyer-ter-Vehn (2001)
based on a numerical simulation (PIC) and cross sections for electron-positron productions, 
in which two opposite laser beams are focussed on a thin gold foil leading to a chemically
non-equiliibrated plasma (see section VI). However, this interesting proposal
still waits for its experimental confirmation. Other production mechanisms, based on the Schwinger
pair production effect in strong fields (Schwinger, 1951), have shown that pair production can be 
efficient already far below the Schwinger critical field strength, requiring laser intensities of 
$5 \times 10^{29}$ W/cm$^2$, in the case of time dependent and inhomogeneous fields, e.g. two
oppositely directed pulsed laser beams in vacuum (see e.g. Avetissian {\it et al.}, 2002;
Narozhny {\it et al.}, 2004; Di Piazza, 2004; Dunne and Schubert, 2005; Gies and Klingmuller, 2005; 
Blaschke {\it et al.} 2006a, Blaschke {\it et al.}, 2006b,
Sch\"utzhold {\it et al.}, 2008). Furthermore the pair production in an X-ray free electron laser 
(XFEL) has been discussed (Ringwald, 2001; Alkofer {\it et al.}, 2001; Roberts {\it et al.}, 2002).
QED plasmas in strong magnetic fields have been considered by Danielsson and Grasso (1995) and 
more general in strong electromagnetic fields by Morozov {\it et al.} (2002).
QED plasmas have also been studied in great detail in the book by Melrose (2008).  

\section{Equation of state}

In this section we will consider the equation of state of an EPP under the following assumptions:

1. ultrarelativistic EPP, i.e. $T \gg m$, 

2. thermal and chemical equilibrium,

3. equal electron and positron density, i.e. vanishing chemical potential,

4. ideal gas, i.e. no interactions in plasma,

5. infinitely extended, homogeneous, and isotropic EPP.

We will relax some of these assumptions in the following sections.
According to these assumptions the distribution function of the electrons and positrons
is given by the Fermi-Dirac distribution
\begin{equation}
n_F (p)=\frac{1}{e^{p/T}+1}
\label{e3}
\end{equation}     
and of the photons by the Bose-Einstein distribution
\begin{equation}
n_F (p)=\frac{1}{e^{p/T}-1},
\label{e4}
\end{equation}     
where the momentum $p$ is identical to the energy $E$ of the particles in the ultrarelativistic
case. It should be noted that the photons are in equilibrium with electrons and positrons under 
the above assumptions, i.e. the system is actually an electron-positron-photon gas.

The particle and energy density can be calculated by integrating over the distribution
functions. In the case of a QGP the results can be found in M\"uller (1985).  
The particle number density of the electrons and positrons follows from integrating over
the Fermi-Dirac distribution as
\begin{equation}
\rho_e^{eq}=g_F \int \frac{d^3p}{(2\pi )^3} n_F(p)=\frac{3}{\pi ^2}\> \zeta (3)\> T^3=0.37\> T^3
\label{e5}
\end{equation}     
where $g_F=4$ is the number of degrees of freedom corresponding to the electrons
and positrons in the two spin states. Assuming a temperature of $T=10$ MeV, we find
$\rho_e^{eq}=370$ MeV$^3 = 4.9 \times 10^{40}$ m$^{-3}$.

The photon density follows accordingly by integrating over the Bose-Einstein distribution with
$g_B=2$ degrees of freedom corresponding to the two polarization states as 
$\rho_{ph}^{eq}=(2/\pi ^2)\> \zeta (3)\> T ^3=0.24\> T^3$.

The energy density of the electron-positron-photon gas is obtained from
\begin{equation}
\epsilon^{eq} = g_F \int \frac{d^3p}{(2\pi )^3} p\, n_F(p) + 
g_B \int \frac{d^3p}{(2\pi )^3} p\, n_B(p) = \frac{11\pi ^2}{60}\> T^4= 1.81\> T^4,
\label{e6}
\end{equation}     
where the photons contribute 36 \% to the energy density. Here the Boltzmann law, 
$\epsilon^{eq}\sim T^4$, holds also for the fermions because we neglected their masses.

For $T=10$ MeV we
find $\epsilon^{eq}= 3.8 \times 10^{29}$ J m$^{-3}$. In a volume of $10^{12}$ m$^3$ 
(corresponding to the size of a neutron star) the total thermal energy of the EPP is
$3.8\times 10^{41}$ J, which corresponds 
to about 10\% of the entire energy (without neutrinos) released in a supernova type II explosion.
In a volume of 1 $\mu$m$^3$ there is still an energy of $3.8 \times 10^{11}$ J contained.

The Coulomb coupling parameter of the EPP, which is a measure for the non-ideal behavior
of a plasma (Ichimaru, 1982), is given by $\Lambda = e^2/(dT)$, where $d\simeq {\rho _e^{eq}}^{-1/3}=
2.7 \times 10^{-14}$ m is the interparticle distance. For $T=10$ MeV we find
$\Lambda = 5.3\times 10^{-3}$ which shows that the EPP is a weakly coupled plasma
in contrast to the QGP where $\Lambda = O(1)$ (Thoma, 2005). Therefore the ideal gas results for
the equation of state derived above are a good approximation. After all, interactions
in the EPP play an important role, for example, for the collective behavior of the
plasma as discussed in the next section and for equilibration of the plasma. Obviously, 
the interaction can be treated by perturbation theory.

\section{Collective Phenomena}

Interactions between particles can be separated into two classes: individual collisions 
between the particles and long-range interactions of particles with the medium (Lifshitz 
and Pitaevskii, 1981).
The latter ones lead to collective effects, which are characteristic for plasmas. The crucial 
quantity from which the collective phenomena are derived is the dielectric tensor 
relating the macroscopic electric field $D_i$ in the medium to the external field $E_i$ 
($i=x,y,z$), i.e. in momentum space
\begin{equation}
D_i(\omega ,{\bf k}) = \sum_j \epsilon_{ij}(\omega ,{\bf k})\> E_j(\omega ,{\bf k}).
\label{e7}
\end{equation}      
In the case of an isotropic medium it depends only on $k=|{\bf k}|$ and has two independent components
\begin{equation}
\epsilon_{ij}(\omega,k) = \epsilon_T(\omega ,k)\> \left (\delta_{ij}-\frac{k_ik_j}{k^2}\right )
+ \epsilon_L(\omega ,k)\> \frac{k_ik_j}{k^2}
\label{e8}
\end{equation}     

The dielectric tensor is closely related to the polarization tensor or photon self-energy by
(see e.g. Elze and Heinz (1989))

\begin{figure}
\centerline{\psfig{figure=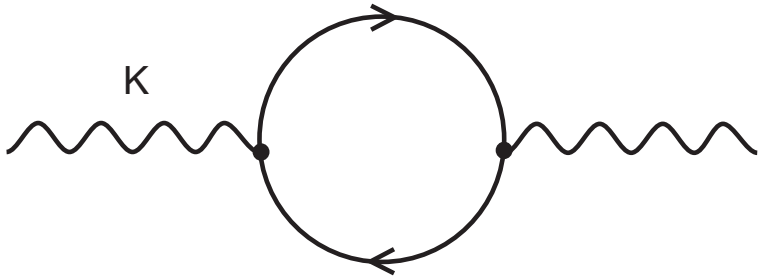,width=10cm}}
\caption{One-loop polarization tensor.}
\end{figure}

\begin{eqnarray}
\epsilon_L(\omega,k) &=& 1-\frac{\Pi_L (\omega,k)}{k^2}\nonumber \\
\epsilon_T(\omega,k) &=& 1-\frac{\Pi_T (\omega,k)}{\omega^2},
\label{e9}
\end{eqnarray}
where $\Pi_L$ and $\Pi_T$ are the longitudinal and transverse components of the polarization tensor,
respectively.

The lowest order diagram for the polarization tensor is shown in Fig.1. Assuming the external
momentum to be soft, i.e. $\omega $ and $k$ to be much smaller than $T$ and the internal
loop momenta to be hard (HTL approximation) an analytic 
result can be found using the real or imaginary time formalism (Klimov, 1982; Weldon, 1982a)
\begin{eqnarray}
\Pi_L(\omega,k) &=& -3m_{ph}^2\> \left (1-\frac{\omega }{2k}\, \ln \frac{\omega + k}{\omega -k}
\right ),\nonumber \\
\Pi_T(\omega,k) &=& \frac{3}{2}\> m_{ph}^2\> \frac{\omega^2}{k^2}\> \left [1-\left (1-
\frac{k^2}{\omega^2}\right )\frac{\omega }{2k}\, \ln \frac{\omega + k}{\omega -k}
\right ],
\label{e10}
\end{eqnarray}
where $m_{ph} = eT/3$ is called the effective photon mass. For $T=10$ MeV we get 
$m_{ph} = 1$ MeV. 

The hard thermal loop approximation,
which is identical to the high-temperature limit, produces consistently the lowest 
order result in finite temperature QED (Braaten and Pisarski, 1990). The dielectric functions following from 
(\ref{e9}) and (\ref{e10}) can also be derived from the classical Vlasov equation together with 
the Maxwell equations (Silin, 1960), since the high-temperature limit corresponds to the 
classical limit. 

The dispersion relations of collective plasma modes, i.e. propagation
of electromagnetic waves in the plasma, can be found by using the 
Maxwell equation, leading to (see e.g. Carrington {\it et al.} (2004))
\begin{eqnarray}
\epsilon_L(\omega,k) &=& 0,\nonumber \\
\epsilon_T(\omega,k) &=& \frac{k^2}{\omega^2}.
\label{e11}
\end{eqnarray}
Combining (\ref{e9}), (\ref{e10}), and (\ref{e11}) gives the dispersion relations 
$\omega_{L,T}(k)$ of the transverse
as well as longitudinal plasma waves as shown in Fig.2. The longitudinal branch, which 
does not exist in vacuum, is called plasmon as in the case of non-relativistic plasmas.
The transverse branch does not play a role in non-relativistic plasmas but is equally
important as the longitudinal one in relativistic plasmas. Both branches start at
the plasma frequency $\omega_{pl}=\omega_{L,T}(k=0)=m_{ph}$ (Kajantie and Kapusta, 1985). 
Consequently the collective plasma waves have soft momenta of the order $eT$.
At high momenta $k\gg m_{ph}$ the transverse mode approaches the free dispersion 
$\omega_T = k$, corresponding to a real photon in vacuum, whereas the longitudinal
mode disappears, i.e. its spectral strength is exponentially suppressed (Pisarski, 1989).
For $T=10$ MeV we find $\omega_{pl}=1.5 \times 10^{21}$ s$^{-1}$. 

Electromagnetic plasma waves in an ultrarelativistic EPP have also been discussed by
Medvedev (1999). For the plasma frequency the same dependence on the temperature
and coupling constant was found (see eq.(3) in Medvedev (1999)) if one uses
the equilibrium number density for the photons proportional to $T^3$ (see above) there.
However, the numerical pre-factor there is wrong and only one 
branch of plasma waves was discussed.   

Another important quantity which can be derived from the polarization or dielectric tensor 
is the Debye screening length, entering the Yukawa potential of
a heavy, non-relativistic test charge in the EPP. The Debye screening length
is given by the static limit of the
longitudinal component of the polarization tensor $1/\Pi_L(\omega =0)$ (Kajantie and Kapusta, 1985), leading to 
$\lambda_D=1/(\sqrt{3}m_{ph})$, which is $1.1 \times 10^{-13}$ m for $T=10$ MeV. 

Finally from (\ref{e10}) we see that the polarization tensor and
the dielectric function become imaginary for $\omega^2<k^2$, i.e.
below the light cone $\omega =k$, corresponding to Landau damping (Pisarski, 1988). We also
observe that the plasma waves calculated at lowest order perturbation theory
are undamped since they are located at $\omega >k$. This changes, however, at higher orders
where the dispersion relation can intersect the light cone (Carrington {\it et al.}, 2004).
It is interesting to note that all these results are apart from some 
numerical color and flavor factors also valid for collective gluon modes in the QGP in the 
high-temperature limit (Thoma, 1995a). 

\begin{figure}
\centerline{\psfig{figure=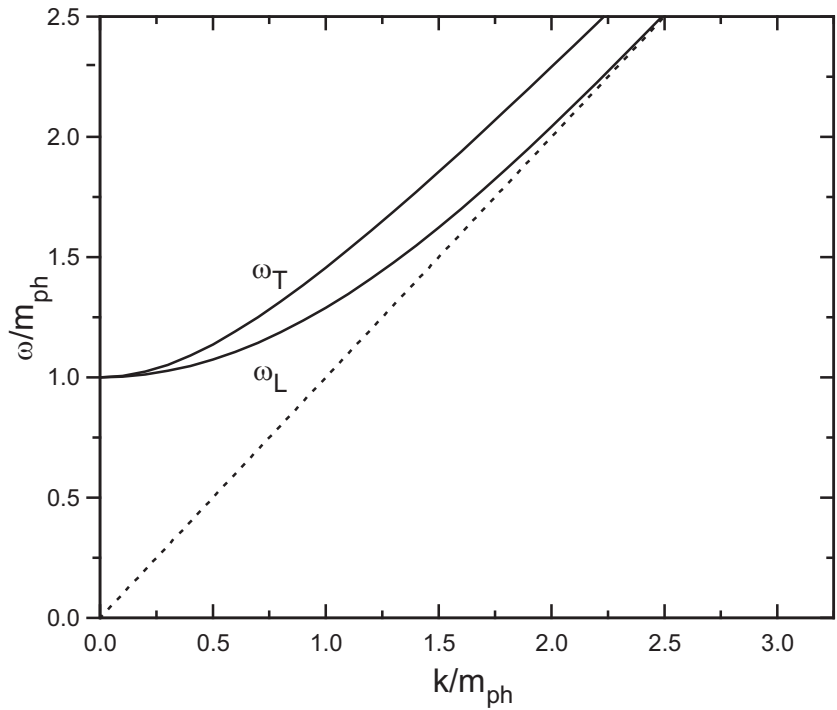,width=11cm}}
\caption{Photon dispersion relation.}
\end{figure}

A complete new phenomenon that does not appear in non-relativistic plasmas is the
existence of fermionic plasma waves because all fermion masses are much too large
in the non-relativistic case. 
Their dispersion relations follow from the pole of the electron
propagator containing the electron self-energy. Using again the hard thermal loop
approximation the electron self-energy reads ($P=(p_0,{\bf p})$, $p=|{\bf p}|$)
(Klimov, 1982; Weldon, 1982b)
\begin{equation}
\Sigma (P) = -a(p_0,p)\> P^\mu \gamma _\mu - b(p_0,p)\> \gamma _0
\label{e13}
\end{equation}
with
\begin{eqnarray}
a(p_0,p) & = & \frac {1}{4p^2}\> \left [tr(P^\mu \gamma _\mu\> \Sigma )
- p_0\> tr(\gamma _0 \> \Sigma )\right ]\; ,\nonumber \\
b(p_0,p) & = & \frac {1}{4p^2}\> \left [P^2\> tr(\gamma _0\> \Sigma )
- p_0\> tr (P^\mu \gamma _\mu \> \Sigma )\right ]\; ,
\label{2.28}
\end{eqnarray}
where the traces over the $\gamma $ matrices are given by
\begin{eqnarray}
tr(P^\mu \gamma _\mu \> \Sigma ) & = & 4\> m_F^2\; ,\nonumber \\
tr(\gamma _0\> \Sigma ) & = & 2\> m_F^2\> \frac {1}{p}\> \ln \frac
{p_0+p}{p_0-p}
\label{e14}
\end{eqnarray}
with the effective electron mass $m_F=eT/\sqrt{8}$ (Klimov, 1982; Weldon, 1982b), which is 1.1 MeV at $T=10$ MeV.

The full electron propagator in the helicity representation is given by (Braaten {\it et al.}, 1990)
\begin{equation}
S^\star (P) = \frac {1}{2D_+(P)}\> (\gamma _0 - \hat p\cdot {\bf \gamma })
            + \frac {1}{2D_-(P)}\> (\gamma _0 + \hat p\cdot {\bf \gamma })
            \; ,
\label{e15}
\end{equation}
where
\begin{equation}
D_\pm (P) = -p_0 \pm p + \frac {1}{4p}\> \left [\pm tr (P^\mu \gamma _\mu
\> \Sigma ) - (\pm p_0 -p)\> tr (\gamma _0\> \Sigma )\right ]\; .
\label{e16}
\end{equation}
Again these results agree with the ones for quarks in the QGP apart from numerical pre-factors. 

The dispersion relations following from the pole of this propagator are shown in Fig.3. 
Again two branches show up, one with a positive ratio of the helicity to chirality ($\omega_+$) following from $D_+=0$,
the other one with a negative ratio ($\omega_-$) following from $D_-=0$, called plasmino (Braaten {\it et al.}, 1990). 
The plasmino branch $\omega_-$, which does not exist in 
vacuum, shows an interesting behavior, namely a minimum at
$k=0.41 m_F$. The same behavior of the quark dispersion in the QGP
has been found. In this case sharp peaks in the dilepton production rate can appear due
to the minimum in the plasmino branch, which leads to van Hove singularities 
(Braaten {\it et al.}, 1990; Peshier and Thoma, 2000).
Whether something similar could be observed in the EPP, e.g. in the electron spectrum, 
is a very interesting question which should be investigated in detail. It could open  
the exciting possibility to observe a new collective plasma wave, the plasmino, 
experimentally in a laser induced EPP.
   
The collective quantities derived here perturbatively from the dielectric functions
are linear phenomena. Nonlinear collective effects in electron-positron plasmas,
have been considered in a number of publications, e.g. soliton formation
(Lontano {\it et al.}, 2001), nonlinear Alfv\'en waves (Zhao {\it et al.}, 1994), nonlinear photon interactions
(Tajima and Taniuti, 1990), plasma enhanced photon splitting (Brodin {\it et al.}, 2007), and nonlinear
self-modulation of radio pulses (Chian and Kennel, 1983). 

\begin{figure}
\centerline{\psfig{figure=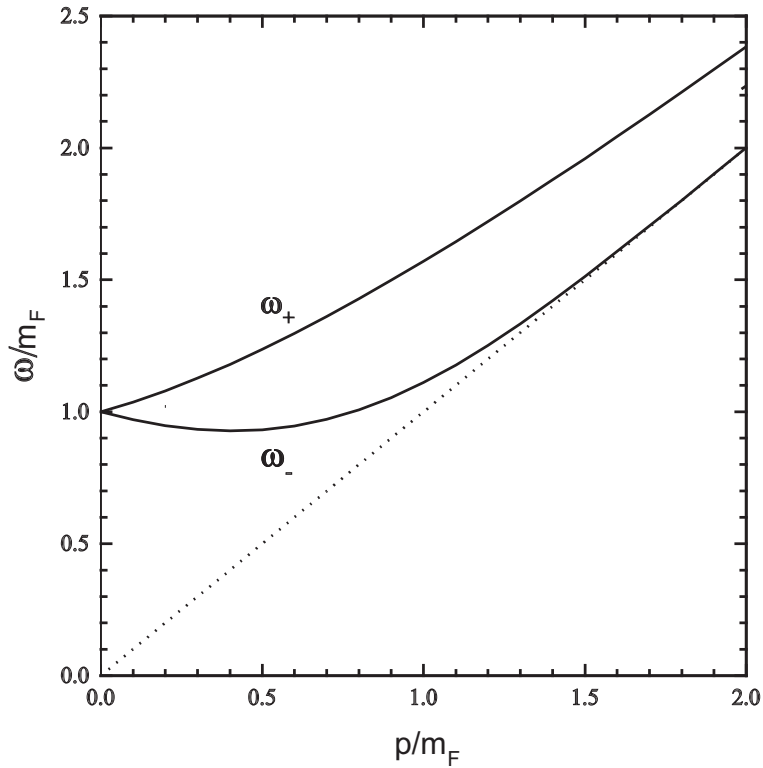,width=10cm}}
\caption{Electron dispersion relation.}
\end{figure}

\section{Transport properties}

Now we want to consider the interaction and properties of particles in the plasma
with hard momenta, i.e. of the order of $T$ or larger. In particular we are interested
in damping and transport rates, mean free paths, collision times, energy losses 
of these particles and other transport properties such as the shear viscosity of the
EPP. All these quantities have been calculated perturbatively in high-temperature QCD
(Thoma, 1995a). As in the case of the collective phenomena, the results can be almost
directly taken from the QCD calculations.

It was shown by Braaten and Pisarski (1990) that a consistent treatment
of gauge theories such as QED at finite temperature, i.e.
for obtaining results that are gauge independent, infrared finite, and complete to 
leading order, require the use of an effective perturbation theory 
using resummed Green functions based on the HTL approximation (HTL resummation technique).  
The HTL method relies on the assumption that one can distinguish between soft and hard 
momenta, i.e. $eT\ll T$. This is approximately fulfilled for QED, where $e=0.3$, but
not in QCD, where the corresponding coupling constant $g>1$.   

The damping rate of an electron or positron in the EPP is defined as the imaginary part of
the dispersion relation $\omega _{L,T}(p)$. To lowest order it follows from the elastic scattering diagram of
Fig.4. In the case of a hard electron or positron with momenta of the order of $T$ or higher 
it exhibits a quadratic infrared (IR)
divergence which can be reduced to a logarithmic one using a HTL resummed photon
propagator. This logarithmic singularity is expected to be cut-off by higher order contributions
leading to (Thoma, 1995a)  
\begin{equation}
\gamma_e = \frac{e^2T}{4\pi}\ln\frac{1}{e}
\label{e17}
\end{equation}     
within logarithmic accuracy, i.e. the constant under the logarithm is not determined.
For $T=10$ MeV we obtain $\gamma_e = 86$ keV, which is much smaller than $\omega_{pl}=1$ MeV,
showing that the EPP is not overdamped. 

\begin{figure}
\centerline{\psfig{figure=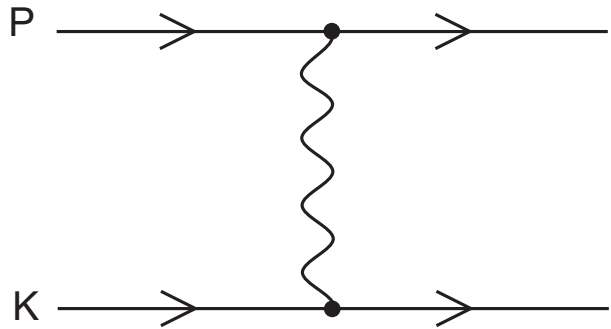,width=8cm}}
\caption{Lowest order diagram for electron-electron scattering.}
\end{figure}

Physically more important are the transport rates $\Gamma $ which are related to the mean free
path and collision time of electrons and positrons in the EPP. They differ from the 
damping rate in cutting off the long range interactions with small scattering
angles $\theta$ by a factor $(1-\cos \theta)$ under the integral defining the
rate (Lifshitz and Pitaevskii, 1981). This leads to an improvement of the IR behavior (logarithmic
instead of quadratic singularity in perturbation theory) and a finite result
using the HTL method. Logarithmic 
divergent quantities can be treated consistently by splitting them into a soft part and
a hard part, where the soft part is calculated using the HTL resummation technique (Braaten and Yuan, 1991).
For the transport rate we find to logarithmic accuracy by extending the QCD
results in Thoma (1994) to QED  
\begin{equation}
\Gamma_e = \frac{e^4 T^3}{3 \pi s}\ln\frac{1}{e},
\label{e18}
\end{equation}     
where the Mandelstam variable $s=(P+K)^2$ is the square of the sum of the four momenta of the 
incoming particles in the scattering diagram of Fig.4. For deriving the above result
we replaced in (15) of Thoma (1994) $C_Ag^2$ by $e^2$ and $m_g$ by $m_{ph}$. For going
beyond the logarithmic approximation, which is valid within about a factor of two, 
one has to calculate explicitly the hard 
contribution from the QED scattering diagrams as it was done by Thoma (1994) for
the QCD case.  
For thermal particles we replace $s$
by its thermal average $\langle s \rangle = 2\langle p\rangle_e \langle k\rangle_e \simeq 19.3 T^2$ 
(Thoma, 1994),
where $\langle p\rangle_e =\langle k\rangle_e = \epsilon_e^{eq}/\rho_e^{eq}= 3.11 T$.
Assuming again $T=10$ MeV, we get $\Gamma_e=0.54$ keV.

The mean free path $\lambda^{mfp}_e$ and collision time $\tau_e$ of the plasma particles (electrons and
positrons)
are given by the inverse of the transport rate $1/\Gamma_e$,
leading to $\lambda^{mfp}_e=0.37$ nm and
$\tau_e = 1.2\times 10^{-18}$ s at $T=10$ MeV.  

In a non-relativistic plasma the shear viscosity can be estimated from elementary kinetic
theory as (Reif, 1965) 
\begin{equation}
\eta _i = \frac {1}{3}\> \sum _i \rho _i\> \langle p_i\rangle \> \lambda^{mfp}_i
\label{e18a}
\end{equation}
where the sum is performed over the various components of the system.
In an relativistic plasma the coefficient 1/3 should be replaced by 4/15 (de Groot {\it et al.}, 1980). Using the mean
free path following from (\ref{e18}), the density of (\ref{e5}), and the thermal momentum $\langle p\rangle_e 
=3.11 T$, 
the shear viscosity is given by (within logarithmic accuracy)
\begin{equation}
\eta_e = \frac{55.8 T^3}{e^4\ln (1/e)}.
\label{e19}
\end{equation}     
At $T=10$ MeV the shear viscosity coefficient is $\eta_e=7.9\times 10^{10}$ Pa s.

Another quantity of interest in a plasma is its stopping power or the energy loss of an energetic
particle in the plasma. There are two contributions, namely the energy loss by collisions
and the radiative one by bremsstrahlung. In a relativistic plasma the latter one becomes important.
The relevance of these contributions in the QGP has been discussed controversially (Mustafa and Thoma, 2005). The collisional 
energy loss is given by the mean energy transfer divided by the mean free path leading to (Braaten and Thoma, 1991)
\begin{equation}
\frac{dE}{dx} = \frac{1}{v} \int d\gamma \> \omega
\label{e19a} 
\end{equation}
where $v$ is the particle velocity, $\gamma$ the damping or interaction rate proportional to the plasma density
and the collision cross section, and $\omega$ the energy transfer from the
energetic particle to plasma particle in the collision.  
Using for the collision cross section the lowest order diagrams in Fig.5, 
the collisional energy of a muon with mass $M$ in an EPP has been calculated 
by Braaten and Thoma (1991) applying the HTL resummation technique 
\begin{equation}
\frac{dE}{dx} = \frac{e^4 T^2}{24 \pi}\>  \left (\frac{1}{v} -
\frac{1-v^2}{2v^2}\> \ln \frac{1+v}{1-v} \right )\>  \left (\ln \frac{E}{M}
+ \ln \frac{1}{e} + A(v) \right ) \;,
\label{e20}
\end{equation}
where $A(v)$ is a slowly varying function of the muon velocity $v$ between 1.3 and 1.5.

\begin{figure}
\centerline{\psfig{figure=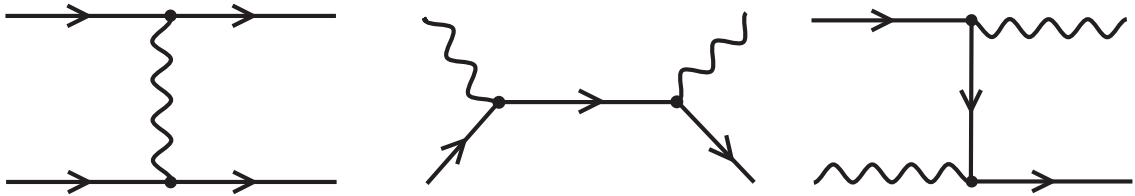,width=15cm}}
\caption{Diagrams defining the collisional energy loss.}
\end{figure}

The collisional energy loss of an electron with energy $E\gg T$ is approximately given by (Braaten and Thoma, 1991)
 \begin{equation}
\frac{dE}{dx}  = \frac{e^4 T^2}{48\pi}\> \ln \frac{15.3E}{e^2T}.
\label{e21}
\end{equation}
In Braaten and Thoma (1991) the factor 15.3 was replaced by 7.6 due to an numerical error which was
corrected in Thoma (1995a). This leads to an energy loss of
200 MeV/nm for an electron or positron with an energy of $E=100$ MeV at $T=10$ MeV,
showing that such an electron is stopped (thermalized) within a fraction of a nanometer.
(Note that the plasma density, proportional to $T^3$, on which the energy loss depends via the mean free path is
hidden in the temperature dependence of the energy loss.) 
Recently it has been shown that this calculation can be improved by taking into
account additional diagrams changing this result slightly (Peigne and Peshier, 2008).
So far no calculations of the radiative energy loss in an EPP have been performed to our knowledge.

The damping rate of a photon in an EPP follows from the diagram in Fig.6, where a HTL resummed
electron propagator has to be used in case of soft momenta of the exchanged electron (positron).
In contrast to the electron damping rate, the photon rate is infrared finite using the HTL method 
due to the presence of an electron propagator in Fig.6 instead of the photon propagator in
Fig.4. Hence there is no need to cut off the long range interaction introducing a transport
cross section. The result for a photon with energy $E=p$ reads (Thoma, 1995b)
\begin{equation}
\Gamma_{ph} = \frac{e^4 T^2}{64 \pi E}\ln\frac{3.88 E}{e^2T}.
\label{e22}
\end{equation}     
The mean free path and the collision time of photons in an EPP are given by $1/\Gamma_{ph}$.
For a thermal photon with the mean momentum 
$\langle p\rangle_{ph} =\epsilon_{ph}^{eq}/\rho_{ph}^{eq}=2.75 T$ at $T=10$ MeV 
the mean free path $\lambda^{mfp}_{ph}=0.28$ nm
and the collision time $\tau_{ph} =9.4\times 10^{-19}$ s follow. Actually the damping rate
given in (\ref{e22}) is a lower limit as higher order effects will enlarge it. As a matter of fact,
the photon production rate in a QGP, which is the inverse process of the damping rate (Thoma, 1995b), 
was shown to be about a factor of 2 larger taking bremsstrahlung into account (Arnold {\it et al.}, 2001).

\begin{figure}
\centerline{\psfig{figure=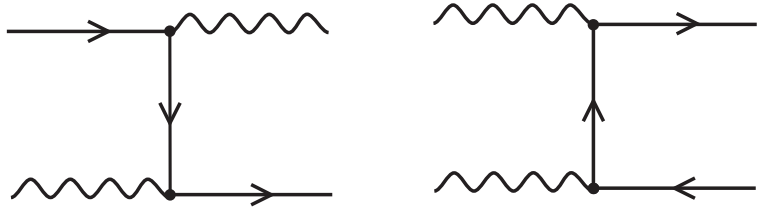,width=10cm}}
\caption{Diagrams defining the photon damping rate.}
\end{figure}

For the viscosity of the photon component using the above mean free path, the
photon density (see above), and the mean photon energy $\langle p\rangle_{ph} = 2.75 T$
we find  
\begin{equation}
\eta_{ph} = \frac{48.7 T^3}{e^4\ln (3.27/e)}
\label{e23}
\end{equation}     
corresponding to $3.5\times 10^{10}$ Pa s at $T=10$ MeV.
Hence the viscosity of the EPP $\eta =\eta_e+\eta_{ph}$ has similar contributions from the
electrons and photons.

A more advanced calculation of the total viscosity of the EPP based on the Kubo formula
yields within logarithmic accuracy (Arnold {\it et al.}, 2000)
\begin{equation}
\eta = \frac{188 T^3}{e^4\ln (1/e)}.
\label{e23a}
\end{equation}
This result is about a factor of 1.5 larger than the one presented here based
on the elementary kinetic theory, which is typically valid within a factor of 2
(Reif, 1965).       

\section{Particle production rates}  

At high temperatures above 10 MeV also other particle species will be
produced, e.g. muons with a mass of $m_\mu = 106$ MeV. Their rate follows 
to lowest order from the diagram in Fig.7 (Born term).
We assume that $m_e \ll T\ll m_\mu$ holds. The first inequality 
implies that the electron mass can be put to zero and the latter inequality 
implies that muons are not equilibrated. Then the muon 
production rate to lowest order ($e^-e^+\rightarrow \gamma^* \rightarrow
\mu^-\mu^+$) is given by 
\begin{equation}
\frac{dN}{d^4xd^4p}=\frac{\alpha ^2}{24\pi^4}\left (1+\frac{2m_\mu^2}{M^2}
\right ) \left (1-\frac{4m_\mu^2}{M^2}\right)^{1/2} \frac{T}{p} 
\frac{1}{\exp(E/T)-1}\ln \frac{1+\exp[-(E+p)/(2T)]}{1+\exp[-(E-p)/(2T)]},
\label{e24}
\end{equation}
where $M^2=E^2-p^2$ is the invariant mass of the virtual photon 
$\gamma ^*$, $E$ its energy and $p=|\bf{p}|$ its momentum. This formula 
was derived by combining the cross section for the process in
Berends {\it et al.} (1973) with the production rate in Cleymans {\it et al.} (1987)
where the quark-antiquark annihilation process 
$q\bar q\rightarrow \mu^-\mu^+$ was considered. Here we assumed the
chemical potential $\mu =0$. The only difference to the process considered 
here comes from the fractional charge of the quarks, i.e. we use here
$\sum_i e_i^2=1$, and from the number of colors in the distribution functions 
in (6a,b) of Cleymans {\it et al.} (1987), i.e. dividing the QGP rate by a factor 
of 9. In addition a factor $3/2$ appears in the electron-positron annihilation 
cross section - see (6) of Berends {\it et al.} (1973) after integrating over the angle -
compared to the quark-antiquark annihilation cross section in (5) of 
Cleymans {\it et al.} (1987), using the relative velocity of the ultrarelativistic 
particles $v_{q\bar q}=1$.
Because of $M^2=E^2-p^2>4 m_\mu^2$ the rate is suppressed exponentially
for temperatures below $2m_\mu$.

In order to obtain the spectrum from this formula one has to integrate over 
the space-time volume, taking into account the space-time evolution 
by using, for example, a hydrodynamical model. The total muon yield then follows from 
integrating the spectrum over the energy and momentum of the virtual photon.

\begin{figure}
\centerline{\psfig{figure=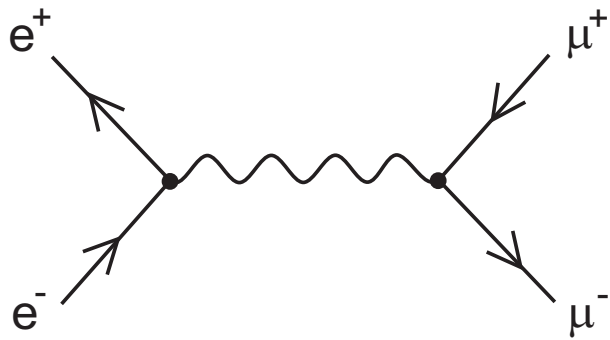,width=8cm}}
\caption{Lowest order contribution of the muon production.}
\end{figure}

At temperatures above 10 MeV also hadron production becomes important, in particular
pion production (Kuznetsova {\it et al.}, 2008). 

\section{Non-equilibrium and Finite Chemical Potential}

EPPs produced in  strong laser fields are probably not in complete equilibrium. For example,
it has been predicted by Shen and Meyer-ter-Vehn (2001) that an positron density of about $5\times 10^{28}$ m$^{-3}$ 
at a temperature of 10 MeV can be reached. This density deviates from the equilibrium
density (\ref{e5}) by 12 orders of magnitude. In the following we will therefore assume that
the EPP produced by lasers is in thermal but not in chemical equilibrium. Then we can replace the 
distribution functions for the electrons by Fermi-Dirac distributions multiplied by
a fugacity factor $\lambda $ describing the deviation from chemical equilibrium, 
$f_F(p)=\lambda n_F(p)$. This assumption has been used
for example for describing the chemical equilibration of the QGP in ultrarelativistic
heavy-ion collisions (Bir\'o {\it et al.}, 1993). The fugacity is given 
by the ratio of the experimental to equilibrium particle density, since the experimental
density follows from integrating over the non-equilibrium distribution, i.e.
\begin{equation}
\rho_{exp}=g_F \int \frac{d^3p}{(2\pi )^3} \lambda n_F(p)=\lambda \rho_{eq} \Rightarrow \lambda = 10^{-12}
\label{e25}
\end{equation}     

Using the real time formalism, QED perturbation theory and the HTL method can also be extended to 
non-equilibrium situations like the one discussed above (Carrington {\it et al.}, 1999). For example, the effective photon mass
is given now by 
\begin{equation}
m_{ph}^2=\frac{4e^2}{3\pi^2} \int_0^\infty dp\> p\> f_F(p)
\label{e26}
\end{equation}     

For $T=10$ MeV we then find for the non-equilibrium photon mass 
$m_{ph}^{noneq} = \sqrt{\lambda} m_{ph} = 1$ eV and the plasma frequency
$\omega_{pl}^{noneq}=1.5 \times 10^{15}$ Hz.
The Debye screening length in such an EPP is $\lambda_D=0.1$ $\mu$m. In order to speak of
a plasma with collective behavior its dimension $L$ should be much larger than
$\lambda_D$, i.e. at least of the order of 1 $\mu$m.

Furthermore, an anisotropic EPP can also be described by quantum field theoretic methods
(Mr\'owczy\'nski and Thoma, 2000). In this case instabilities can occur (Mr\'owczy\'nski and Thoma, 2007).

Finally a possible difference between the positron density and the electron density can be 
treated by introducing a finite chemical potential $\mu$, i.e. using the distribution
\begin{equation}
n_F (p)=\frac{1}{e^{(p\pm \mu)/T}+1}
\label{e27}
\end{equation}     
for the electrons (negative sign) and positrons (positive sign).
Such a difference comes from the fact that the laser produced EPP is embeded in a hot electron
and cold ion background of the target. Therefore there will be an excess of electrons over positrons
in the hot, relativistic EPP.  
The methods described above,
such as the HTL resummation, can be generalized easily to this case (Vija and Thoma, 1995). For example,
the energy density is given by
\begin{equation}
\epsilon^{eq} = \frac{11\pi ^2}{60}\> T^4 + \frac{1}{2}\> T^2\mu^2+\frac{1}{4\pi^2}\> \mu^4
\label{e28}
\end{equation}   
or the effective photon energy by 
\begin{equation}
m_{ph}^2=\frac{e^2T^2}{9} \left (1+\frac{3\mu^2}{\pi^2 T^2}\right ).
\label{e29}
\end{equation}
The muon production rate for finite chemical potential can be found by replacing
the last three factors in (\ref{e24}) by (13) of Cleymans {\it et al.} (1987). 

\section{Conclusions}

The properties of ultrarelativistic EPPs can be treated by QED perturbation
theory at finite temperature similarly 
as the QGP by QCD using the imaginary or real time formalism. 
A consistent treatment requires the use of the HTL resummation technique for many
quantities under consideration.
Many results can be more or less directly taken from the corresponding QGP calculation.
The examples discussed here are the equation of state, the dielectric tensor,
the dispersion relations of photons and electrons, the Debye screening length,
damping and transport rates, mean free paths, collision times, shear viscosity,
stopping power, and the muon production rate. In particular collective phenomena,
such as plasma waves, are of interest. In an EPP completely new phenomena 
appear, namely collective fermion modes associated with the possibility
to observe van Hove singularities. 

EPPs produced by strong lasers open the unique possibility to investigate 
these plasmas and their properties which also exist in astrophysical 
systems, e.g. supernova explosions. However, current predictions indicate
that laser induced EPPs are not in chemical equilibrium. Their properties therefore
require an extension of perturbative QED and the HTL resummation technique
to non-equilibrium situations, which is possible within the real time formalism.
As a first application we discussed the Debye screening length in such an EPP, 
demonstrating that the EPP should have a spatial extension of at least one micron 
to exhibit a characteristic plasma behavior.

Here we have considered
a number of relevant properties of an ultrarelativistic, 
equilibrated EPP as a reference for future laboratory experiments 
and astrophysical EPPs
and gave numbers for a plasma temperature of 10 MeV. The results are
summarized in the table below.

New results, which have not been published so far, are the electron transport rate
(\ref{e18}), the electron viscosity (\ref{e20}), and the photon viscosity (\ref{e24}).
Also new is the application of the results, derived here or compiled from the literature,
to relativistic laser plasmas as presented in the third row of the table and in the discussion
of non-equilibrium effects. 

Further quantities, e.g. the radiative energy loss, or higher
order corrections, e.g. beyond the leading logarithm approximation,
to the quantities considered above could be evaluated
in a similar way as discussed above. Also an extension of the properties discussed in sections
II - V to the case of a chemical non-equilibrated EPP would be of interest.
Finally we did not consider here the formation of an EPP in a strong
laser field, e.g. by the Schwinger mechanism, and its equilibration.

\medskip

{\bf Acknowledgment:} I would like to thank D. Habs, J. Rafelski, I. Kouznetsova, and G. Moore
for helpful discussions and hints.

\newpage


\noindent
\leftline{Table: Properties of an EPP}

\smallskip

\noindent
\small
\begin{tabular}{l|l|p{4cm}}
Quantity& Formula& Value at $T=10$ MeV\\ \hline
Electron-Positron Density& $\rho_e^{eq}=3/\pi ^2\> \zeta (3)\> T^3$& $4.9 \times 10^{40}$ m$^{-3}$\\
Photon Density& $\rho_{ph}^{eq}=2/\pi ^2\> \zeta (3)\> T^3$& $3.2 \times 10^{40}$ m$^{-3}$\\
Electron-Positron Energy Density& $\epsilon_e^{eq}=7\pi ^2/60\> T^4$& $2.4 \times 10^{29}$ J m$^{-3}$\\
Photon Energy Density& $\epsilon_{ph}^{eq}=\pi ^2/15\> T^4$& $1.4 \times 10^{29}$ J m$^{-3}$\\
Total Energy Density& $\epsilon^{eq}=11\pi ^2/60\> T^4$& $3.8 \times 10^{29}$ J m$^{-3}$\\
Thermal electron momentum& $\langle p\rangle_e = \epsilon_e^{eq}/\rho_e^{eq}=3.11\> T$& 31 MeV\\
Thermal photon momentum& $\langle p\rangle_{ph} = \epsilon_{ph}^{eq}/\rho_{ph}^{eq}=2.75\> T$& 28 MeV\\
Interparticle distance& $d\simeq {\rho_e^{eq}}^{-1/3}$& $2.7\times 10^{-14}$ m\\
Coulomb Coupling Parameter& $\Lambda = e^2/(dT)$& $5.3\times 10^{-3}$\\
Effective Photon Mass& $m_{ph}=eT/3$& 1 MeV\\
Plasma Frequency& $\omega_{pl}=m_{ph}$& $1.5 \times 10^{21}$ s$^{-1}$\\
Debye Screening Length& $\lambda_D=1/(\sqrt{3}m_{ph})$& $1.1 \times 10^{-13}$ m\\
Effective Electron Mass& $m_F=eT/(2\sqrt{2})$& 1.1 MeV\\
Electron Damping Rate& $\gamma_e = e^2T/(4\pi)\> \ln (1/e)$& 86 keV\\
Electron Transport Rate& $\Gamma_e = e^4 T^3/(3 \pi s)\> \ln (1/e)$& 0.54 keV for $s=19.3\> T^2$\\
Photon Damping Rate& $\Gamma_{ph} = e^4 T^2/(64 \pi E)\> \ln (3.88 E/e^2T)$& 0.70 keV for $E=2.75\> T$\\
Electron Mean Free Path& $\lambda_e^{mfp}=1/\Gamma_e$& 0.37 nm\\
Photon Mean Free Path& $\lambda^{mfp}_{ph}=1/\Gamma_{ph}$& $0.28$ nm\\
Electron Collision Time& $\tau_e=1/\Gamma_e$& $1.2\times 10^{-18}$ s\\ 
Photon Collision Time& $\tau_{ph} =1/\Gamma_{ph}$& $9.4\times 10^{-19}$ s\\
Electron Viscosity& $\eta_e = 55.8\> T^3/[e^4\ln(1/e)]$& $7.9\times 10^{10}$ Pa s\\
Photon Viscosity& $\eta_{ph} = 48.7\> T^3/[e^4\ln (3.27/e)]$& $3.5\times 10^{10}$ Pa s\\
Total Viscosity& $\eta =\eta_e+\eta_{ph}$& $(1.1 - 1.6)\times 10^{11}$ Pa s\\
Electron Energy Loss& $dE/dx=e^4T^2/(48\pi )\> \ln (15.3E/e^2T)$& 200 MeV/nm for $E=100$ MeV\\
\end{tabular}

\normalsize

\bigskip

\setlength{\parindent}{0pt}

{\bf References}

\bigskip

Alkofer, R., M. B. Hecht, C. D. Roberts, S. M. Schmidt, and D. V. Vinnik, 2001, Phys. Rev. Lett. {\bf 87}, 193902.  

Altherr, T., and U. Kraemmer, 1992, Astropart. Phys. {\bf 1}, 133. 

Arnold, P., L. G. Yaffe, and G. D. Moore, 2000, JHEP {\bf 0011}, 001.

Arnold, P., G. D. Moore, and L. Yaffe, 2001, JHEP {\bf 0112}, 009.

Avetissian, H. K., A. K. Avetissian, G. F. Mkrtchian, and K. V. Sedrakian, 2002, Phys. Rev. E {\bf 66}, 016502.

Berends, F. A., K. J. F. Gaemers, and R. Gastmans, 1973, Nucl. Phys. B {\bf 57}, 381.

Beskin, V. S., A. V. Gurevich, and Y. N. Istomin, 1993, {\it Physics of the Pulsar Magnetosphere}, 
(Cambridge University Press, Cambridge).

Bir\'o, T. S., E. van Doorn, B. M\"uller, M. H. Thoma, and X. N. Wang, 1993,
Phys. Rev. C {\bf 48}, 1275.

Blaschke, D. B., A.V. Prozorkevich, C. D. Roberts, S. M. Schmidt, and S. A. Smolyansky, 2006a, 
Phys. Rev. Lett. {\bf 96}, 140402. 
 
Blaschke, D. B., A.V. Prozorkevich, S. A. Smolyansky, and A. V. Tarakanov, 2006b, J. Phys: Conf. Ser. {\bf 35}, 121.

Braaten, E., and R. D. Pisarski, 1990,  Nucl. Phys. B {\bf 337}, 569.

Braaten, E., R. D. Pisarski, and T. C. Yuan, 1990, Phys. Rev. Lett. {\bf 64}, 2242.

Braaten, E., and M. H. Thoma, 1991, Phys. Rev. D {\bf 44}, 1298.

Braaten, E., and T. C. Yuan, 1991, Phys. Rev. Lett. {\bf 66}, 2183.

Brodin, G., M. Marklund, B. Eliasson, and P. K. Shukla, 2007, Phys. Rev. Lett. {\bf 98}, 125001.  

Carrington, M. E., H. Defu, and M. H. Thoma, 1999, Eur. Phys. J. C {\bf 67},
347.

Carrington, M. E., T. Fugleberg, D. Pickering, and M. H. Thoma, 2004,
Can. J. Phys. {\bf 82}, 671.

Chian, A. C.-L., and C. F. Kennel, 1983, Astrophys. Space Sci. {\bf 97}, 9.

Cleymans, J., J. Fingberg, and K. Redlich, 1987, Phys. Rev. D {\bf 35}, 2153.

Danielsson, U. H., and D. Grasso, 1995, Phys. Rev. D {\bf 52}, 2533.

Di Piazza, A., 2004, Phys. Rev. D {\bf 70}, 053013.

Dunne, G. V., and C. Schubert, 2005, Phys. Rev D {\bf 72}, 105004.

Elze, H. T., and U. Heinz, 1989, Phys. Rep. {\bf 183}, 81.

Gies, H., and K. Klingmuller, 2005, Phys. Rev. D {\bf 72}, 065001.

de Groot, S. R., W. A. van Leeuwen, and C. G. van Weert, 1980,
{\it Relativistic Kinetic Theory} (North-Holland, Amsterdam).

Gyulassy, M. and L. McLerran, 2005, Nucl. Phys. A {\bf 750}, 30. 


Hardy, S. J., and M. H. Thoma, 2001, Phys. Rev. D {\bf 63}, 025014.

Ichimaru, S., 1982, Rev. Mod. Phys. {\bf 54}, 1017.

Kajantie, K., and J. I. Kapusta, 1985, Ann. Phys. (N.Y.) {\bf 160}, 477.

Kapusta, J. I., 1989, {\it Finite Temperature Field Theory}
(Cambridge University Press, New York).

Klimov, V. V.,  1982, Zh. Eksp. Teor. Fiz. {\bf 82}, 336 [Sov. Phys. JETP {\bf 55}, 199 (1982)].

Kuznetsova, I., D. Habs, and J Rafelski, 2008, Phys. Rev. D {\bf 78}, 014027.

Landsmann, N. P., and C. G. van Weert, 1987, Phys. Rep. {\bf 145}, 141. 

Le Bellac, M., 1996, {\it Thermal Field Theory} (Cambridge University Press, Cambridge).

Liang, E. P., S. C. Wilks, and M. Tabak, 1998, Phys. Rev. Lett. {\bf 81}, 4887.

Lifshitz, E. M., and L. P. Pitaevskii, 1981, {\it Physical Kinetics} 
(Pergamon Press, Oxford).

Lontano, M., S. Bulanov, and J. Koga, 2001, Phys. Plasmas {\bf 8}, 5113.

Marklund, M., and P. K. Shukla, 2006, Rev. Mod. Phys. {\bf 78}, 591.

Matsubara, T., 1955, Progr. Theor. Phys. {\bf 14}, 351.

Melrose, D. B., 2008, {\it Quantum Plasmaphysics - Unmagnetized Plasmas} (Springer, Berlin).

Medvedev, M. V., 1999, Phys. Rev. E {\bf 59}, R4766.

Morozov, V. G., G. R\"opke, and A. H\"oll, 2002, Theor. Math. Phys. {\bf 131}, 812; ibid. {\bf 132}, 1029. 

Mourou, G. A., T. Tajima, and S. V. Bulanov, 206, Rev. Mod. Phys. {\bf 78}, 309. 

Mr\'owczy\'nski, S., and M. H. Thoma, 2000, Phys. Rev. D {\bf 62},
036011.

Mr\'owczy\'nski, S., and M. H. Thoma, 2007, Annu. Rev. Nucl. Part.
Sci. {\bf 57}, 61.

Mustafa, M. G., and M. H. Thoma, 2005, Acta Phys. Hung. A {\bf 22}, 93.

M\"uller, B., 1985, {\it The Physics of the Quark-Gluon Plasma},
Lecture Notes in Physics 225 (Springer, Berlin).

Narozhny, N. B., S. S. Bulanov, V. D. Mur, and V. S. Popov, 2004, Phys. Lett. A {\bf 330}, 1. 

Peigne, S. and A. Peshier, 2008, Phys. Rev. D {\bf 77}, 014015.

Peshier, A., and M. H. Thoma, 2000, Phys. Rev. Lett. {\bf 84}, 841.

Pisarski, R. D., 1988, Nucl. Phys. B {\bf 309}, 476.

Pisarski, R. D., 1989, Physica A {\bf 158}, 146.

Raffelt, G., 1996, {\it Stars as Laboratories for Fundamental Physics}, (Univ. Chicago Press, Chicago).

Reif, F., 1965, {\it Fundamentals of Statistical and Thermal Physics}, (McGraw-Hill, New York).

Ringwald, A., 2001, Phys. Lett. B {\bf 510}, 107.

Roberts, C. D., S. M. Schmidt, and D. V. Vinnik, 2002, Phys. Rev. Lett. {\bf 89}, 153901.

Sch\"utzhold, R., Gies H., and G. Dunne, eprint arXiv:0807.0754.

Schwinger, J., 1951, Phys. Rev. {\bf 82}, 664. 

Shen, B., and J. Meyer-ter-Vehn, 2001, Phys. Rev. E {\bf 65}, 016405.

Silin, V. P., 1960, J. Exptl. Theoret. Phys. (U.S.S.R.) {\bf 38}, 1577 [Sov. Phys. JETP {\bf 11}, 1136 (1960)].

Tajima, T. and T. Taniuti, 1990, Phys. Rev. A {\bf 42}, 3587.

Thoma, M. H., 1994, Phys. Rev. D {\bf 49}, 451.

Thoma, M. H., 1995a, in {\it Quark-Gluon Plasma 2}, edited by R.C. Hwa
(World Scientific, Singapore, 1995), p.51 [eprint hep-ph/9503400].

Thoma, M. H. 1995b, Phys. Rev D {\bf 51}, 862.

Thoma, M. H., 2002, in Space Science Series of ISSI, Vol.14,
edited by P. Jetzer, K. Pretzl, and R. von Steiger (Kluwer Academic Publishers,
Dordrecht), p.141 [eprint astro-ph/0104078].

Thoma, M. H., 2005, J. Phys. G {\bf 31}, L7 and Erratum J. Phys. G {\bf 31}, 539. 

Tsytovich, V. N., 1961, J. Exptl. Theoret. Phys. (U.S.S.R.) {\bf 40}, 1775 [Sov. Phys. JETP {\bf 13}, 1249 (1961)].

Vija, H., and M. H. Thoma, 1995, Phys. Lett. B {\bf 342}, 212.

Weldon, H. A., 1982a, Phys. Rev. D {\bf 26}, 1394.

Weldon, H. A., 1982b, Phys. Rev. D {\bf 26}, 2789.

Zel'dovich, Y. B., and Novikov, 1971, {\it Relativistic Astrophysics, Vol. 1: Stars and Relativity}, 
(Univ. Chicago Press, Chicago).

Zel'dovich, Y. B., and Novikov, 1983, {\it Relativistic Astrophysics, Vol. 2: the Structure and Evolution of the Universe}, 
(Univ. Chicago Press, Chicago).

Zhao, J., K. I. Nishikawa, J. I. Sakai, and T. Neubert, 1994, Phys. Plasmas {\bf 1}, 103. 

\end{document}